\documentclass[aps,prl,superscriptaddress,showpacs,twocolumn,floatfix,nofootinbib]{revtex4-2}

\usepackage{graphicx,colordvi}
\usepackage{color}
\usepackage{csvsimple}
\usepackage{rotating}
\usepackage{lineno} 
\usepackage{amsmath}

\usepackage[]{hyperref}

\begin{document}
\title{First Measurement of the EMC Effect in $^{10}$B and $^{11}$B}

\newcommand{\UVA }{University of Virginia, Charlottesville, Virginia 22903, USA}
\newcommand{\REG }{University of Regina, Regina, Saskatchewan S4S 0A2, Canada}
\newcommand{\NCAT }{North Carolina A \& T State University, Greensboro, North Carolina 27411, USA}
\newcommand{\KENT }{Kent State University, Kent, Ohio 44240, USA}
\newcommand{\ZAG }{University of Zagreb, Zagreb, Croatia}
\newcommand{\TEMP }{Temple University, Philadelphia, Pennsylvania 19122, USA}
\newcommand{\YER }{A.I. Alikhanyan  National  Science  Laboratory \\ (Yerevan  Physics Institute),  Yerevan  0036,  Armenia}
\newcommand{\MSU }{Mississippi State University, Mississippi State, Mississippi 39762, USA}
\newcommand{\WM }{The College of William \& Mary, Williamsburg, Virginia 23185, USA}
\newcommand{\CUA }{Catholic University of America, Washington, DC 20064, USA}
\newcommand{\HU }{Hampton University, Hampton, Virginia 23669, USA}
\newcommand{\FIU }{Florida International University, University Park, Florida 33199, USA}
\newcommand{\CNU }{Christopher Newport University, Newport News, Virginia 23606, USA}
\newcommand{\JAZ }{Jazan University, Jazan 45142, Saudi Arabia}
\newcommand{\JLAB }{Thomas Jefferson National Accelerator Facility, Newport News, Virginia 23606, USA}
\newcommand{\UTENN }{University of Tennessee, Knoxville, Tennessee 37996, USA}
\newcommand{\OHIO }{Ohio University, Athens, Ohio 45701, USA}
\newcommand{\UCONN }{University of Connecticut, Storrs, Connecticut 06269, USA}
\newcommand{\SBU }{Stony Brook University, Stony Brook, New York 11794, USA}
\newcommand{\ODU }{Old Dominion University, Norfolk, Virginia 23529, USA}
\newcommand{\ANL }{Argonne National Laboratory, Lemont, Illinois 60439, USA}
\newcommand{\BOULDER }{University of Colorado Boulder, Boulder, Colorado 80309, USA}
\newcommand{\ORSAY }{Institut de Physique Nucleaire, Orsay, France}
\newcommand{\UNH }{University of New Hampshire, Durham, New Hampshire 03824, USA}
\newcommand{\JMU }{James Madison University, Harrisonburg, Virginia 22807, USA}
\newcommand{\RUTG }{Rutgers University, New Brunswick, New Jersey 08854, USA}
\newcommand{\CMU }{Carnegie Mellon University, Pittsburgh, Pennsylvania 15213, USA}
\newcommand{\LBL}{Lawrence Berkeley National Laboratory, Berkeley, California 94720, USA}
\newcommand{\VUNION}{Department of Natural Sciences, Virginia Union University, Richmond, Virginia 23220, USA}
\newcommand{\VTECH}{Present address: Virginia Tech, Blacksburg, Virginia 24061, USA}
\newcommand{\SHANDONG}{Present address: Shandong University, Qingdao, Shandong 266237, China}
\newcommand{\ALZ}{Present address: Physics Department,  Al-Zaytoonah University of Jordan, Amman 11733, Jordan}

\author{A.~Karki}\affiliation{\MSU}           
\author{D.~Biswas}\thanks{\VTECH}\affiliation{\HU}
\author{F.~A.~Gonzalez}\affiliation{\SBU}      
\author{W.~Henry}\affiliation{\JLAB}
\author{C.~Morean}\affiliation{\UTENN}
\author{A.~Nadeeshani}\affiliation{\HU}
\author{A.~Sun}\affiliation{\CMU}
\author{D.~Abrams}\affiliation{\UVA}
\author{Z.~Ahmed}\affiliation{\REG} 
\author{B.~Aljawrneh}\thanks{\ALZ}\affiliation{\NCAT} 
\author{S.~Alsalmi}\affiliation{\KENT}
\author{R.~Ambrose}\affiliation{\REG}
\author{D.~Androic}\affiliation{\ZAG} 
\author{W.~Armstrong}\affiliation{\ANL}
\author{J.~Arrington}\affiliation{\LBL}
\author{A.~Asaturyan}\affiliation{\YER}
\author{K.~Assumin-Gyimah}\affiliation{\MSU}
\author{C.~Ayerbe Gayoso}\affiliation{\WM}\affiliation{\MSU}   
\author{A.~Bandari}\affiliation{\WM}  
\author{J.~Bane}\affiliation{\UTENN}  
\author{J.~Barrow}\affiliation{\UTENN}
\author{S.~Basnet} \affiliation{\REG} 
\author{V.~Berdnikov}\affiliation{\CUA}
\author{H.~Bhatt}\affiliation{\MSU}  
\author{D. Bhetuwal}\affiliation{\MSU}
\author{W.~U.~Boeglin}\affiliation{\FIU}
\author{P.~Bosted}\affiliation{\WM}
\author{E.~Brash}\affiliation{\CNU}
\author{M.~H.~S.~Bukhari}\affiliation{\JAZ}
\author{H.~Chen}\affiliation{\UVA}       
\author{J.~P.~Chen}\affiliation{\JLAB}  
\author{M.~Chen}\affiliation{\UVA}     
\author{M.~E.~Christy}\affiliation{\HU} 
\author{S.~Covrig}\affiliation{\JLAB}  
\author{K.~Craycraft}\affiliation{\UTENN} 
\author{S.~Danagoulian}\affiliation{\NCAT} 
\author{D.~Day}\affiliation{\UVA}         
\author{M.~Diefenthaler}\affiliation{\JLAB}
\author{M.~Dlamini}\affiliation{\OHIO}     
\author{J.~Dunne}\affiliation{\MSU}        
\author{B.~Duran}\affiliation{\TEMP}       
\author{D. Dutta}\affiliation{\MSU}
\author{C.~Elliott}\affiliation{\UTENN} 
\author{R.~Ent}\affiliation{\JLAB} 
\author{H.~Fenker}\affiliation{\JLAB}
\author{N.~Fomin}\affiliation{\UTENN} 
\author{E.~Fuchey}\affiliation{\UCONN} 
\author{D.~Gaskell}\affiliation{\JLAB}  
\author{T.~N.~Gautam}\affiliation{\HU}  
\author{J.~O.~Hansen}\affiliation{\JLAB}         
\author{F.~Hauenstein}\affiliation{\ODU}      
\author{A.~V.~Hernandez}\affiliation{\CUA}   
\author{T.~Horn}\affiliation{\CUA}          
\author{G.~M.~Huber}\affiliation{\REG}      
\author{M.~K.~Jones}\affiliation{\JLAB}       
\author{S.~Joosten}\affiliation{\ANL}        
\author{M.~L.~Kabir}\affiliation{\MSU}
\author{N.~Kalantarians}\affiliation{\VUNION}  
\author{C.~Keppel}\affiliation{\JLAB}        
\author{A.~Khanal}\affiliation{\FIU}        
\author{P.~M.~King}\affiliation{\OHIO}      
\author{E.~Kinney}\affiliation{\BOULDER}    
\author{H.~S.~Ko}\affiliation{\ORSAY}       
\author{M.~Kohl}\affiliation{\HU}         
\author{N.~Lashley-Colthirst}\affiliation{\HU} 
\author{S.~Li}\affiliation{\UNH}           
\author{W.~B.~Li}\affiliation{\WM}         
\author{A.~H.~Liyanage}\affiliation{\HU}   
\author{D.~Mack}\affiliation{\JLAB}        
\author{S.~Malace}\affiliation{\JLAB}      
\author{P.~Markowitz}\affiliation{\FIU}   
\author{J.~Matter}\affiliation{\UVA}     
\author{D.~Meekins}\affiliation{\JLAB}   
\author{R.~Michaels}\affiliation{\JLAB}  
\author{A.~Mkrtchyan}\affiliation{\YER}  
\author{H.~Mkrtchyan}\affiliation{\YER}  
\author{S.~Nanda}\affiliation{\MSU} 
\author{D.~Nguyen}\affiliation{\UVA}
\author{G.~Niculescu}\affiliation{\JMU} 
\author{I.~Niculescu}\affiliation{\JMU} 
\author{Nuruzzaman}\affiliation{\RUTG}  
\author{B.~Pandey}\affiliation{\HU}     
\author{S.~Park}\affiliation{\SBU}      
\author{E.~Pooser}\affiliation{\JLAB}   
\author{A.~J.~R.~Puckett}\affiliation{\UCONN}  
\author{M.~Rehfuss}\affiliation{\TEMP}     
\author{J.~Reinhold}\affiliation{\FIU}     
\author{N.~Santiesteban}\affiliation{\UNH} 
\author{B.~Sawatzky}\affiliation{\JLAB}    
\author{G.~R.~Smith}\affiliation{\JLAB}    
\author{H. Szumila-Vance}\affiliation{\JLAB}
\author{A.~S.~Tadepalli}\affiliation{\RUTG}    
\author{V.~Tadevosyan}\affiliation{\YER}     
\author{R.~Trotta}\affiliation{\CUA}        
\author{S.~A.~Wood}\affiliation{\JLAB}     
\author{C.~Yero}\affiliation{\FIU} 
\author{J.~Zhang}\thanks{\SHANDONG}\affiliation{\SBU}     
\collaboration{for the Hall C Collaboration}
\noaffiliation

\noaffiliation 

\date{\today}

\begin{abstract}

The nuclear dependence of the inclusive inelastic electron scattering cross section (the EMC effect) has been measured for the first time in $^{10}$B and $^{11}$B.  Previous measurements of the EMC effect in $A \leq 12$ nuclei showed an unexpected nuclear dependence;  $^{10}$B and $^{11}$B were measured to explore the EMC effect in this region in more detail.  Results are presented for $^9$Be, $^{10}$B, $^{11}$B, and $^{12}$C at an incident beam energy of 10.6~GeV.  The EMC effect in the boron isotopes was found to be similar to that for $^9$Be and $^{12}$C, yielding almost no nuclear dependence in the EMC effect in the range $A=4-12$.  This represents important, new data supporting the hypothesis that the EMC effect depends primarily on the local nuclear environment due to the cluster structure of these nuclei.

\end{abstract}

\maketitle


\section{Introduction}
Deep inelastic electron scattering from nuclear targets provides access to the inelastic structure functions, which are connected to the quark distributions (parton distribution functions) in the nucleus. The modification of structure functions in nuclei (the EMC effect) is a clear indication that the nucleus cannot be simply described in terms of on-shell nucleon degrees of freedom. Despite intense theoretical and experimental study since its first observation in 1983~\cite{EuropeanMuon:1983wih}, there remain multiple theoretical explanations of the origin of the EMC effect~\cite{Malace:2014uea, Cloet:2019mql}.

The observation that the EMC effect appears to scale with local (rather than average) nuclear density~\cite{Seely:2009gt} instigated a paradigm shift in possible explanations of the effect. In Ref.~\cite{Seely:2009gt}, it was found that the size of the EMC effect for $^3$He, $^4$He and $^{12}$C appeared to scale well with average nuclear density.  However, the EMC effect in $^9$Be was similar in size to $^4$He and $^{12}$C, despite having a significantly smaller average density.  Since the beryllium nucleus can be described as two $\alpha$ particles with a single neutron, it was hypothesized that the EMC effect is driven by the density of nucleons in those clusters (local nuclear density). It was subsequently found that the relative number of short-range correlated nucleon pairs (SRCs) in a nucleus (inferred from the ratio of the inclusive electron scattering cross section at $x>1$ between nuclei and the deuteron) exhibited a similar density dependence~\cite{Fomin:2011ng}. Additional studies directly examined the correlation of the size of the EMC effect with SRCs~\cite{Hen:2012fm, Arrington:2012ax}.  The high degree of correlation between these two nuclear effects reinforces the idea that the local nuclear environment plays an important role in the EMC effect. One explanation posits that the EMC effect is driven by changes in the nucleon structure due to local changes in nuclear density~\cite{Arrington:2012ax}. It has also been suggested that the apparent connection between the EMC effect and SRCs can come about from highly virtual nucleons in a correlated pair, leading to large off-shell effects~\cite{Weinstein:2010rt}. Within the precision of existing data, both explanations have been found to be consistent with the observed correlation between the EMC effect and SRCs~\cite{Arrington:2012ax, CLAS:2019vsb, Arrington:2019wky}.

The local density (LD) and high virtuality (HV) hypotheses can be further explored by making additional measurements of the EMC effect and SRC ratios. More data on light nuclei will improve our understanding of the underlying nuclear physics driving both SRCs and the EMC effect. In addition, measurements at nearly-constant values of $A$ covering a range in $N/Z$  will help us understand the impact of the isospin structure (since SRCs are dominated by neutron-proton pairs~\cite{Tang:2002ww, Subedi:2008zz, Arrington:2011ax, fomin17, Arrington:2022sov}).  Such measurements will be made at Jefferson Lab in experimental Hall C by experiments E12-10-008 (EMC) and E12-06-105 (SRC)~\cite{E1210008,E1206105}. As part of the group of commissioning experiments that ran in Hall C after the completion of the Jefferson Lab 12 GeV Upgrade, a small subset of the planned EMC data were taken.  We report on the results from this commissioning run, extracting the first measurement of the EMC effect in $^{10}$B and $^{11}$B.  The boron isotopes are of interest due to the fact that, like $^9$Be, they are also expected to have significant $\alpha$ cluster contributions to their nuclear structure, while at the same time have an average density noticeably different from both $^9$Be and $^{12}$C.  Measurement of the EMC effect in $^{10,11}$B could provide additional confirmation that, as noted in Ref.~\cite{Seely:2009gt}, the $\alpha$ cluster configuration (and hence local nuclear density) plays a significant role or, alternately, indicate that $^9$Be is an outlier for other reasons yet to be determined. 

\section{Experimental Details and Analysis}
This experiment ran in parallel with JLab E12-10-002 (a measurement of inclusive electron scattering from hydrogen and deuterium) for about two days in February, 2018. The electron beam with energy 10.602$\pm$0.004~GeV impinged on 10~cm long liquid hydrogen (LH2) and liquid deuterium (LD2) cryogenic targets and several solid targets: $^9$Be, $^{12}$C, $^{10}$B$_4$C, and $^{11}$B$_4$C.  The B$_4$C targets were isotopically enriched to (at least) 95\% by weight. The contribution from carbon to the B$_4$C yield was subtracted using measured yields from the carbon target.

Scattered electrons were detected in the new Super High Momentum Spectrometer (SHMS), a superconducting magnetic focusing spectrometer in a QQQD (three quadrupoles followed by a single dipole) configuration, with an additional small dipole (3$^\circ$ horizontal bend) just before the first quadrupole to allow access to small scattering angles. The SHMS has a nominal solid angle of $\approx4.0$~msr with a fractional momentum acceptance of $-10\%<\frac{\Delta P}{P_0}<22\%$. 

A detector package after the final dipole was used to identify electrons and provide tracking information for angle and momentum reconstruction.  This detector package includes a pair of horizontal drift chambers, each chamber containing six planes of wires oriented at 0$^\circ$ and $\pm$60$^\circ$ with respect to horizontal. The drift chambers provided position and direction information at the spectrometer focal plane; momentum and angle information  at the target were reconstructed from this information via a fitted matrix transformation.  The detector hut also includes four hodoscope planes (three planes of scintillators and one quartz bar plane) for triggering and timing.  The hodoscopes are also used to help determine the tracking efficiency (typically 95-96\%) by using a subset of paddles to define a region through which events were sure to have traversed the drift chambers.  A gas Cherenkov (filled with 1 atm of CO$_2$) and a lead-glass calorimeter were used for electron identification.  The event trigger required the presence of hits in three of the four hodoscope planes as well as the presence of a signal in either the gas Cherenkov or calorimeter. Due the high efficiency of the hodoscopes and the conservative thresholds used in the event trigger, the trigger efficiency was better than 99.9\%. The detector package also includes another gas Cherenkov (typically filled with C$_4$F$_8$O at pressures below 1 atm) and an aerogel detector; these last two detectors were present in the detector stack and active but were not used in the analysis of data from this experiment as they are primarily used for separation of pions, kaons, and protons rather than electron identification.  

Additional measurements at the same central angle but over a reduced kinematic range were also made in the High Momentum Spectrometer (HMS).  Since the HMS was used extensively in the Jefferson Lab 6~GeV program, its performance and acceptance are more thoroughly understood than those of the SHMS and was used as a systematic check of the resulting target cross section ratios.

For the results presented in this work, measurements were made at a single SHMS central angle (21$^\circ$) and three central momentum settings; $P_0=3.3$, 4.0, and 5.1 GeV.  These spectrometer settings resulted in a coverage in Bjorken $x$ of 0.3 to 0.95, while the negative of the four-momentum transfer squared, $Q^2$, varied from 4.3 to 8.3~GeV$^2$.  The invariant mass of the hadronic system, $W$, is larger than 2~GeV (i.e. above the nominal nucleon resonance region) up to $x\approx0.7$.

Electron yields were binned in the fractional spectrometer momentum ($\Delta P/P_0$) and corrected for detector and tracking efficiencies as well as computer and electronic deadtimes. An additional correction was applied to the cryogenic targets for target density reduction due to beam heating. Backgrounds to the electron yields included pion contamination and contributions from charge symmetric processes.  The latter were measured directly by flipping the spectrometer polarity and measuring the resulting positron yields.  The positron yields scaled approximately with the radiation length of the target and were at most $\approx$1\% of the electron yield at negative polarity.  The pion contamination was determined by examining calorimeter spectra in the region where the electron signal is expected to dominate, selecting pions using the gas Cherenkov, and was at most 0.5\% at low $x$. For values of $x$ at which the pions were above threshold in the gas Cherenkov detector ($x=0.58$), the pion contamination grew to be as large as 1.2\%.  For the cryotargets, contribution to the yield from the aluminum walls of the target cells was measured using two aluminum foils at the same positions along the beam as the ends of the cryotarget.  The contribution to the yield was measured to be about 5\% of the LD2 target yield with little variation as a function of $x$.  As noted earlier, the contribution from carbon to the B$_4$C target yield was measured using the $^{12}$C target.  This contribution was about 20\% of the B4C target yield.  Since the shape of the carbon distribution is very similar to that from the subtraction B$_4$C target, the resulting cross section ratios were relatively insensitive to the size of the carbon contribution.

Yields were converted to cross sections via the Monte Carlo ratio method:
\begin{equation}
  \left(\frac{d\sigma}{d\Omega dE'}\right)_{\textrm{exp}} = \frac{Y_{\textrm{exp}}}{Y_{\textrm{sim}}} \left(\frac{d\sigma}{d\Omega dE'}\right)_{\textrm{model}},
\end{equation}
where $Y_{\textrm{exp}}$ is the efficiency corrected, background subtracted experimental yield, $Y_{\textrm{sim}}$ is the Monte Carlo yield produced using a model cross section, radiated using the Mo and Tsai formalism~\cite{Mo:1968cg,Tsai:1971qi,Dasu:1993vk}, and $\left(\frac{d\sigma}{d\Omega dE'}\right)_{\textrm{model}}$ is the same model used to produce the simulated yield evaluated at Born level. The model cross section uses a fit~\cite{Bosted:2012qc} based on a superscaling~\cite{Maieron:2001it} approach for the quasielastic  contribution. The inelastic cross section is based on a fit to the inelastic deuteron structure function~\cite{Bodek:1979rx} modified by a fit to the EMC effect~\cite{Gomez:1993ri} for $W^2>3.0$~GeV$^2$, while for $W^2<2.0$~GeV$^2$, the cross section is calculated using a convolution over the nucleon structure function (similar to that described in~\cite{Arrington:2021vuu}). The region $2.0$~GeV$^2<W^2<3.0$~GeV$^2$ is taken as the weighted average between the low $W^2$ and high $W^2$ calculations.  The sensitivity of the extracted $\sigma_A/\sigma_D$ ratios to the cross section model used in this analysis was tested by using alternate fits to the quasielastic and inelastic cross sections (in particular, the model described in~\cite{Arrington:2021vuu} and a new parameterization based on a global fit to world data~\cite{christy_priv}) and was found to be small (typically on the order of 0.4\%). Target cross section ratios were formed for each ($\Delta P/P_0$) bin, converted to $x$, and grouped in bins of fixed width in $x$, ($\Delta x =0.025$).

\begin{figure*}[htb]
{\includegraphics*[width=16cm]{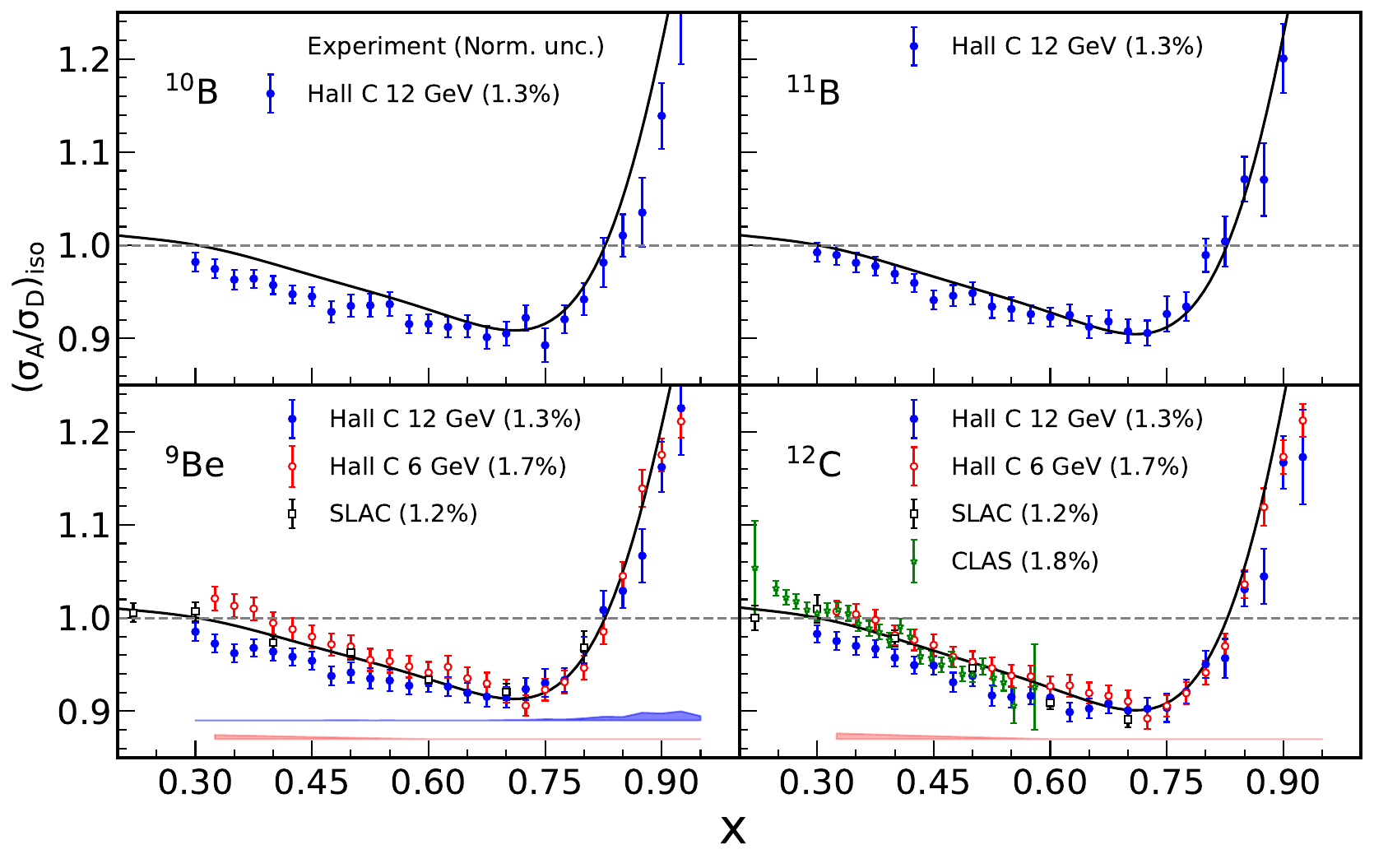}}
\caption{Ratio of isoscalar-corrected cross section per nucleon vs. $x$, for $^9$Be, $^{10}$B, $^{11}$B, and $^{12}$C from this experiment (blue, closed circles). The $^9$Be and $^{12}$C plots include the final results from JLab Hall C at 6 GeV~\cite{Arrington:2021vuu} (open red circles) as well as those from SLAC E139~\cite{Gomez:1993ri} (open black squares).  Also shown are the carbon results from JLab CLAS at 6 GeV~\cite{CLAS:2019vsb} (green stars).  Error bars include statistics combined in quadrature with point-to-point systematic errors while the normalization error for each experiment is noted in the label.  The red band denotes the $x$-correlated error for the JLab Hall C 6~GeV results, while the blue band shows the $x$-correlated error for this experiment (only shown for beryllium since it is largely target independent).  The solid black curve is the $A$-dependent fit of the EMC effect from SLAC E139~\cite{Gomez:1993ri}.
\label{emc_vs_x}}
\end{figure*}

In addition to the typical radiative and acceptance corrections applied in the extraction of cross sections, two additional corrections were used when determining the $\sigma_A/\sigma_D$ cross-section ratios. First, so-called isoscalar corrections were applied to $^9$Be and $^{11}$B to account for the difference between the inelastic neutron and proton cross sections, $\sigma_n$ and $\sigma_p$:
\begin{equation}
  \left(\frac{\sigma_A}{\sigma_D}\right)_{\textrm{ISO}}
  = \frac{\frac{A}{2}(\sigma_p+\sigma_n)}{(Z\sigma_p+N\sigma_n)} \frac{\sigma_A}{\sigma_D}
  = \frac{\frac{A}{2}(1+\frac{\sigma_n}{\sigma_p})}{(Z+N\frac{\sigma_n}{\sigma_p})} \frac{\sigma_A}{\sigma_D},
\end{equation}
where $A$ and $Z$ are the atomic weight and atomic number, with $N=A-Z$, and $\sigma_A/\sigma_D$ is the cross section ratio per nucleon. As described in Ref.~\cite{Arrington:2021vuu}, we use the effective cross sections for nucleons bound in the deuteron~\cite{Arrington:2011qt} to evaluate $\sigma_n/\sigma_p$.  An additional correction is also applied to account for acceleration (deceleration) of the incoming (outgoing) electrons in the Coulomb field of the nucleus.  This correction is calculated using a modified version of the Effective Momentum Approximation (EMA)~\cite{Aste:2004yz, Seely:2009gt} and in the DIS region ranges from 0.16\% at $x=0.3$ to 0.5\% at $x=0.7$ for carbon (smaller for lighter nuclei).  The correction increases at larger $x$,  reaching $\approx$0.8\% at $x=0.95$.

We divided the systematic uncertainty in the EMC cross section ratios into three categories: point-to-point, $x$-correlated, and normalization uncertainties.  Note that some quantities can contribute to more than one kind of uncertainty.
\begin{itemize}
\item Point-to-point uncertainties are assumed to be independent for each target and $x$-bin and contribute to the uncertainty in a manner similar to the statistical uncertainty.  The largest of these uncertainties include those assigned to account for variation in the beam current/charge calibration over time (0.34\%), variations across the spectrometer momentum bite in the extended target acceptance as compared to the thin, solid targets (0.5\%), and kinematic dependent contributions to the radiative corrections (0.5\%).  Other, smaller contributions included those from electronic dead time, detector efficiency, and target density reduction.  The total point-to-point uncertainty in the EMC ratios was estimated to be 0.87\%.

\item So-called $x$-correlated uncertainties vary in size with $x$, but impact all points simultaneously.  These include uncertainties due primarily to kinematic quantities, like beam energy, scattering angle, and spectrometer central momentum.  In the region $x$=0.3-0.7, these uncertainties are on the order of 0.1\%, but can grow to 1.22\% at the very largest values of $x$. 

\item Normalization uncertainties contribute to all points collectively, affecting the overall scale of the ratio.  Significant sources of normalization uncertainty include the LD2 target thickness (0.6\%) and density reduction due to target boiling (0.3\%), LD2 target wall subtraction (0.5\%), solid target thicknesses (0.5-0.66\%), and a contribution to the radiative correction uncertainty due to the difference in target radiation lengths and input cross-section models (0.5\%).  An additional 0.5\% normalization uncertainty was assigned to account for possible acceptance issues hypothesized to explain the difference in EMC ratios observed between the SHMS and HMS.  The total normalization uncertainty was 1.22-1.29\%.

Note that when comparing the $\sigma_A/\sigma_D$ ratios, the contribution to the normalization uncertainty from the LD2 target thickness and associated target boiling (0.68\%) and the LD2 target wall subtraction (0.5\%) are common to all targets and should be removed when comparing, e.g., $^{12}$C to $^{10}$B.
\end{itemize}

\section{Results and Discussion}
The EMC ratios as a function of $x$ for all four nuclei measured in this experiment ($^9$Be, $^{11}$B, $^{10}$B, and $^{12}$C) are shown in Figure~\ref{emc_vs_x}. Our results for $^9$Be and $^{12}$C are plotted along with those from the JLab Hall C 6 GeV experiment~\cite{Seely:2009gt} and SLAC E139~\cite{Gomez:1993ri}.  Results from the CLAS spectrometer in Hall B at 6~GeV~\cite{CLAS:2019vsb} are also shown for carbon. The $A$-dependent fit of the EMC effect from SLAC E139~\cite{Gomez:1993ri} is shown (solid black curve) for each ratio. In general, there is reasonable agreement between data sets for $^9$Be and $^{12}$C with respect to the $x$ dependence of the ratio. The ratios for $^{10}$B and $^{11}$B are the first measurement of the EMC effect for these nuclei.  Numerical values for the EMC ratios shown in Fig.~\ref{emc_vs_x} are available on request.

Upon extraction of the EMC ratios shown in Figure~\ref{emc_vs_x}, it was found that the C and Be results were systematically smaller than previous measurements by about 2\% with a significance of 2~$\sigma$. Subsequent investigation found no issues with the data analysis that would impact the ratio. Cross-checks with data taken in the HMS over a more limited $x$ range showed some disagreement (at the 0.5\% level) with the SHMS, suggesting there were effects due to differing acceptance for long 10~cm targets compared to the much shorter solid targets, but not large enough to explain the entire discrepancy.  We hypothesize that there may be an unknown effect with respect to the deuterium target thickness or density. In the interpretation of the data, we focus on the slope of the EMC ratio between $0.3<x<0.7$ as a primary measurement of the size of the EMC effect. The impact of a possible 2\% normalization offset is small compared to the size of the relative uncertainties of the extracted slopes (which are on the order of 12\%) so has minimal impact on the interpretation of the results.

In addition to the overall normalization issue described above, there is some tension between $^9$Be results for this measurement and the Hall C 6 GeV measurement at low $x$ that merits some discussion. The kinematics of the Hall C 6 GeV data (low momentum and large scattering angle) resulted in a large contribution from the radiated quasi-elastic tail at low $x$.  This, combined with the relatively large radiation length of the Be target made the 6 GeV data very sensitive to the model used to determine the radiated quasi-elastic cross section.  It is possible the systematic uncertainty was underestimated. In contrast, the radiated quasielastic tail contribution is much smaller for the data presented here.

\begin{figure}[htb]
\begin{center}
\includegraphics[width=\columnwidth]{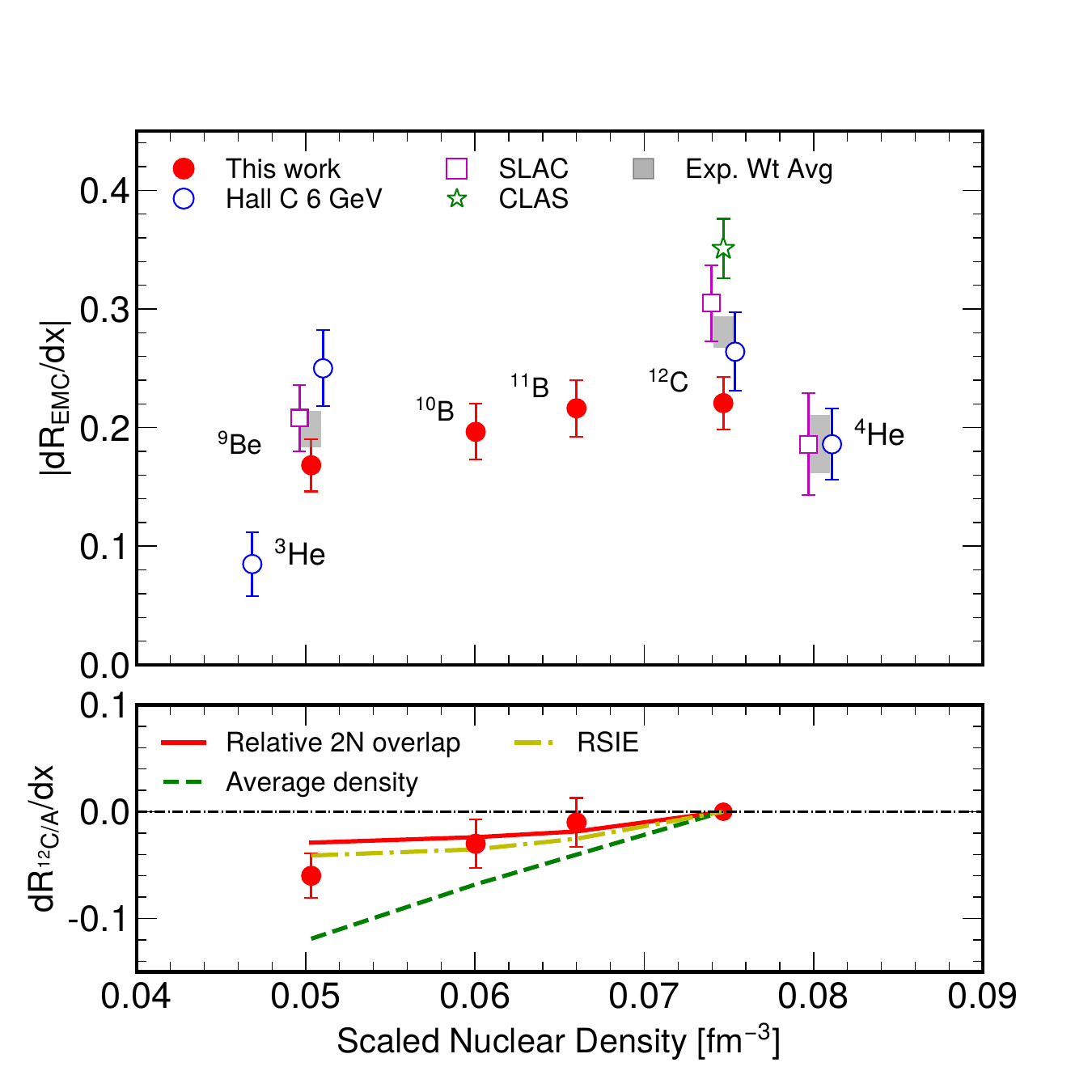}
\caption{Top: Size of the EMC effect (slope from the cross section ratio for $0.3<x<0.7$) vs. scaled nuclear density ($\rho (A-1)/A)$ for $^3$He, $^4$He, $^9$Be, $^{10,11}$B, and $^{12}$C. Closed circles are from this work, open circles from the JLab Hall C 6~GeV results~\cite{Arrington:2021vuu}, open squares from SLAC E139~\cite{Gomez:1993ri}, and the open star from CLAS at 6~GeV~\cite{CLAS:2019vsb}.  Some points have been offset horizontally for visibility. Grey bands denote the weighted average of all experiments shown for a given target (where applicable). Bottom: Slope extracted from the cross section ratios of $^{12}$C to $^9$Be, $^{12}$C to $^{10}$B, and $^{12}$C to $^{11}$B from this experiment. The red and blue curves are calculations of the EMC effect assuming scaling with relative 2N overlap or average nuclear density (see text).  The yellow curve is from a calculation of the EMC effect based on the residual strong interaction (RSIE)~\cite{Wang:2014rxa}. All calculations have been normalized to the slope for carbon.
\label{emc_vs_density}}
\end{center}
\end{figure}

The size of the EMC effect can be more precisely described using the magnitude of the slope, $|dR_{\textrm{EMC}}/dx|$ in the region $0.3<x<0.7$ (the ``EMC region''). These slopes are shown in Figure~\ref{emc_vs_density} (top), where the magnitude of the EMC effect is plotted vs.\ the scaled nuclear density. The scaled nuclear density is calculated from Green's Function Monte Carlo calculations of the nucleon spatial distributions~\cite{Pieper:2001mp} with a correction (slightly reducing the effective density) applied to account for the finite size of the nucleon. As in Ref.~\cite{Seely:2009gt}, the density is scaled by $(A-1)/A$ to account for the fact that we are interested in the effect of the $A-1$ nucleons on the struck nucleon.  Note that the densities presented here are slightly different from those in Ref.~\cite{Seely:2009gt}, due primarily to updated calculations for carbon, resulting most visibly in a change in the relative density as compared to $^4$He (previously, the resulting density for carbon was larger than that for $^4$He).  The EMC slopes from this experiment include an additional systematic uncertainty of 0.009 ($\approx$ 4.5\% of the slope) from the fact that, although the slope was fit over a fixed range in $x$, variations in that choice of $x$ interval lead to changes in the extracted slope. 

Fig.~\ref{emc_vs_density} (top) also includes slopes from all experimental results included in Fig.~\ref{emc_vs_x}. Grey bands denote the combination of all experiments for a given target, where applicable. With the higher precision provided by this determination of the size of the EMC effect, some tension between the data sets is apparent. For $^9$Be, the 6 GeV Hall C data and the results from this work are both in agreement with the SLAC E139 results, but are in some disagreement with each other.  This is likely due to systematic effects from the cross section model used in the radiative corrections which are larger for the 6 GeV data (as discussed earlier).  On the other hand, the 6 GeV Hall C results agree with those from this experiment for carbon, although the latter are in some tension with the SLAC E139 and CLAS ratios. It is also worth noting that the EMC ratios from the CLAS experiment for all targets (in addition to $^{12}$C, the CLAS results include $^{27}$Al, $^{56}$Fe, and $^{208}$Pb) are systematically larger than those from other experiments, as discussed in Ref.~\cite{Arrington:2021vuu}.  It is possible that the systematic difference in the CLAS results can be attributed to differences in the approximations used in determination of the radiative corrections as compared to those from the SLAC and Hall C experiments.

We can more precisely compare the size of the EMC effect in $^{12}$C to the other targets studied in this experiment by taking the direct cross section ratio of $^{12}$C to $^9$Be, $^{10}$B, and $^{11}$B (see Fig.~\ref{emc_vs_density}, bottom plot).  By taking the ratio between solid targets directly, the statistical uncertainty from deuterium is eliminated and the systematic errors are slightly smaller.  The slight difference between $^{12}$C and $^{9}$Be ($3.2\sigma$) and $^{10}$B ($1.4\sigma$) is now apparent.

\begin{figure}[htb]
\begin{center}
\includegraphics[width=\columnwidth]{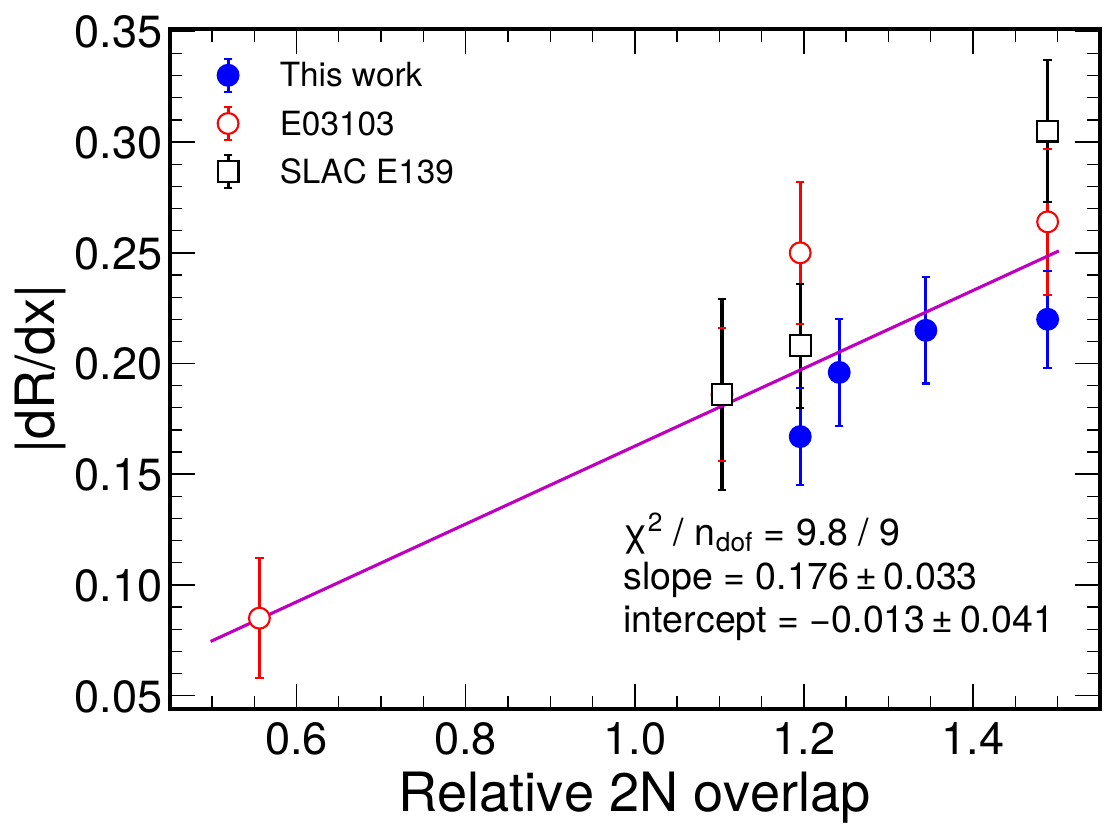}
\caption{Size of the EMC effect vs. relative 2N overlap, $\langle O_N \rangle-\langle O_D\rangle$. Data points are the same as in Fig.~\ref{emc_vs_density}.  The datum from CLAS is excluded due to inconsistencies with world data as well as possible systematic effects from the use of a different approach to radiative corrections.}
\label{emc_vs_overlap}
\end{center}
\end{figure}

The $^{12}$C/A ratios are also compared to three predictions for the nuclear dependence of the EMC effect.  While the three models discussed here can provide information about the origins of the EMC effect via examination of the nuclear dependence, none of these models provide predictions for the absolute magnitude of the EMC effect. The first describes the EMC effect in terms of the residual strong interaction energy~\cite{Wang:2014rxa}.  The residual strong interaction energy (RSIE) is a refinement of the nuclear binding energy, corrected for Coulomb contributions: $RSIE(A,Z) = B(A,Z) + a_c Z (Z-1) A^{-1/3}$, where $B(A,Z)$ is the nuclear binding energy (given by the Bethe-Weizs{\"a}cker formula~\cite{Samanta:2002sp,heyde1999basic}) and the constant $a_c$ in the Coulomb contribution term is 0.71~MeV. The second prediction assumes the EMC effect scales with average nuclear density, with the constraint that the EMC effect is zero for the deuteron.  The third calculation assumes that the EMC effect is driven by the relative two-nucleon (2N) overlap in the nucleus, $\langle O_N \rangle-\langle O_D\rangle$~\cite{Arrington:2012ax}.  The relative 2N overlap is calculated using two-nucleon distributions from GFMC calculations~\cite{Pieper:2001mp} to estimate the relative probability to find two nucleons within a certain distance. A direct comparison of the EMC effect vs. relative 2N overlap is shown in Fig.~\ref{emc_vs_overlap} (note that the values of relative 2N overlap in this pot correspond to the 1.7~fm hard-cutoff version in Ref.~\cite{Arrington:2012ax}, red triangles in Fig. 9 of that reference).  There is clearly an excellent correlation between the two quantities. The slope and intercept from a linear fit to all the data shown in Fig.~\ref{emc_vs_overlap} are consistent with a fit that includes only prior data (slope=0.216 $\pm$ 0.038, intercept=-0.039 $\pm$ 0.044) indicating that these new results (which add $^{10}$B and $^{11}$B) support the dependence observed earlier.

\begin{table}[htb]
    \centering
    \begin{tabular}{|c|c|c|} \hline
    Target     & $|dR_{\textrm{EMC}}/dx|$ & $dR_{^{12}\textrm{C}/A}/dx$  \\ \hline
    $^9$Be     & 0.168 $\pm$ 0.022       & -0.060 $\pm$ 0.019 \\
    $^{10}$B   & 0.196 $\pm$ 0.024       & -0.030 $\pm$ 0.021 \\
    $^{11}$B   & 0.216 $\pm$ 0.024       & -0.010 $\pm$ 0.021 \\
    $^{12}$C   & 0.221 $\pm$ 0.022       & --                 \\ \hline
    \end{tabular}
    \caption{Slopes of EMC ratios extracted in this work.  The second column shows the slopes from the $A/D$ ratios while the last column gives the ratios of $^{12}$C/$A$ to more precisely study the relative EMC effect in $^{9}$Be, $^{10}$B, $^{11}$B, and $^{12}$C.}
    \label{tab:emc_slopes}
\end{table}

The results shown in Fig.~\ref{emc_vs_density} and Tab.~\ref{tab:emc_slopes} suggest that there is little nuclear dependence of the EMC effect for $^4$He, $^9$Be, $^{10}$B, $^{11}$B, and $^{12}$C. While the average of all results for carbon yields a larger EMC effect than the other nuclei, the average would decrease from 0.280$\pm$0.013 to 0.252$\pm$0.016 if the CLAS data were excluded. In Ref.~\cite{Seely:2009gt} it was suggested that the relatively large EMC effect in $^9$Be could be explained by its $\alpha$ cluster structure and the idea that the EMC effect is driven by local density.  $^{10}$B and $^{11}$B are also thought to have significant $\alpha$ cluster contributions to their nuclear structure~\cite{Zhusupov:2020nkn, Itagaki:2021fpk}, and were chosen for this reason. The similarity of the boron results to $^4$He, $^9$Be, and $^{12}$C serves as confirmation of the $\alpha$ cluster hypothesis and that local nuclear effects play a significant role in the EMC effect.  The correlation between the size of the EMC effect and the relative 2N overlap provides further support for the importance of the local nuclear environment in the EMC effect.

\section{Conclusions}
In summary, we have made the first measurement of the EMC effect in $^{10}$B and $^{11}$B, providing new information on the nuclear dependence of the EMC effect.  The size of the EMC effect for the boron isotopes is similar to that for $^4$He, $^9$Be, and $^{12}$C, reinforcing the hypothesis that the EMC effect is driven by local, rather than average nuclear density.  A clear correlation between the size of the EMC effect and the relative 2N overlap in a  nucleus is observed, giving further support for the importance of the local nuclear properties in the EMC effect. It will be particularly interesting to see if SRC ratios from the boron isotopes follow the same trend as the EMC effect. 

\section{Acknowledgments}
This material is based upon work supported by the U.S. Department of Energy, Office of Science, Office of Nuclear Physics under contracts DE-AC05-06OR23177, DE-AC02-05CH11231, DE-SC0013615, and DE-FE02-96ER40950, and by the Natural Sciences and Engineering Research Council of Canada (NSERC) SAPIN-2021-00026.

\bibliographystyle{apsrev4-1}
\bibliography{emc_boron}

\begin{thebibliography}{34}%
\makeatletter
\providecommand \@ifxundefined [1]{%
 \@ifx{#1\undefined}
}%
\providecommand \@ifnum [1]{%
 \ifnum #1\expandafter \@firstoftwo
 \else \expandafter \@secondoftwo
 \fi
}%
\providecommand \@ifx [1]{%
 \ifx #1\expandafter \@firstoftwo
 \else \expandafter \@secondoftwo
 \fi
}%
\providecommand \natexlab [1]{#1}%
\providecommand \enquote  [1]{``#1''}%
\providecommand \bibnamefont  [1]{#1}%
\providecommand \bibfnamefont [1]{#1}%
\providecommand \citenamefont [1]{#1}%
\providecommand \href@noop [0]{\@secondoftwo}%
\providecommand \href [0]{\begingroup \@sanitize@url \@href}%
\providecommand \@href[1]{\@@startlink{#1}\@@href}%
\providecommand \@@href[1]{\endgroup#1\@@endlink}%
\providecommand \@sanitize@url [0]{\catcode `\\12\catcode `\$12\catcode
  `\&12\catcode `\#12\catcode `\^12\catcode `\_12\catcode `\%12\relax}%
\providecommand \@@startlink[1]{}%
\providecommand \@@endlink[0]{}%
\providecommand \url  [0]{\begingroup\@sanitize@url \@url }%
\providecommand \@url [1]{\endgroup\@href {#1}{\urlprefix }}%
\providecommand \urlprefix  [0]{URL }%
\providecommand \Eprint [0]{\href }%
\providecommand \doibase [0]{http://dx.doi.org/}%
\providecommand \selectlanguage [0]{\@gobble}%
\providecommand \bibinfo  [0]{\@secondoftwo}%
\providecommand \bibfield  [0]{\@secondoftwo}%
\providecommand \translation [1]{[#1]}%
\providecommand \BibitemOpen [0]{}%
\providecommand \bibitemStop [0]{}%
\providecommand \bibitemNoStop [0]{.\EOS\space}%
\providecommand \EOS [0]{\spacefactor3000\relax}%
\providecommand \BibitemShut  [1]{\csname bibitem#1\endcsname}%
\let\auto@bib@innerbib\@empty
\bibitem [{\citenamefont {Aubert}\ \emph {et~al.}(1983)\citenamefont {Aubert}
  \emph {et~al.}}]{EuropeanMuon:1983wih}%
  \BibitemOpen
  \bibfield  {author} {\bibinfo {author} {\bibfnamefont {J.~J.}\ \bibnamefont
  {Aubert}} \emph {et~al.} (\bibinfo {collaboration} {European Muon}),\ }\href
  {\doibase 10.1016/0370-2693(83)90437-9} {\bibfield  {journal} {\bibinfo
  {journal} {Phys. Lett. B}\ }\textbf {\bibinfo {volume} {123}},\ \bibinfo
  {pages} {275} (\bibinfo {year} {1983})}\BibitemShut {NoStop}%
\bibitem [{\citenamefont {Malace}\ \emph {et~al.}(2014)\citenamefont {Malace},
  \citenamefont {Gaskell}, \citenamefont {Higinbotham},\ and\ \citenamefont
  {Cloet}}]{Malace:2014uea}%
  \BibitemOpen
  \bibfield  {author} {\bibinfo {author} {\bibfnamefont {S.}~\bibnamefont
  {Malace}}, \bibinfo {author} {\bibfnamefont {D.}~\bibnamefont {Gaskell}},
  \bibinfo {author} {\bibfnamefont {D.~W.}\ \bibnamefont {Higinbotham}}, \ and\
  \bibinfo {author} {\bibfnamefont {I.}~\bibnamefont {Cloet}},\ }\href
  {\doibase 10.1142/S0218301314300136} {\bibfield  {journal} {\bibinfo
  {journal} {Int. J. Mod. Phys. E}\ }\textbf {\bibinfo {volume} {23}},\
  \bibinfo {pages} {1430013} (\bibinfo {year} {2014})}\BibitemShut {NoStop}%
\bibitem [{\citenamefont {Clo\"et}\ \emph {et~al.}(2019)\citenamefont {Clo\"et}
  \emph {et~al.}}]{Cloet:2019mql}%
  \BibitemOpen
  \bibfield  {author} {\bibinfo {author} {\bibfnamefont {I.~C.}\ \bibnamefont
  {Clo\"et}} \emph {et~al.},\ }\href {\doibase 10.1088/1361-6471/ab2731}
  {\bibfield  {journal} {\bibinfo  {journal} {J. Phys. G}\ }\textbf {\bibinfo
  {volume} {46}},\ \bibinfo {pages} {093001} (\bibinfo {year}
  {2019})}\BibitemShut {NoStop}%
\bibitem [{\citenamefont {Seely}\ \emph {et~al.}(2009)\citenamefont {Seely}
  \emph {et~al.}}]{Seely:2009gt}%
  \BibitemOpen
  \bibfield  {author} {\bibinfo {author} {\bibfnamefont {J.}~\bibnamefont
  {Seely}} \emph {et~al.},\ }\href {\doibase 10.1103/PhysRevLett.103.202301}
  {\bibfield  {journal} {\bibinfo  {journal} {Phys. Rev. Lett.}\ }\textbf
  {\bibinfo {volume} {103}},\ \bibinfo {pages} {202301} (\bibinfo {year}
  {2009})}\BibitemShut {NoStop}%
\bibitem [{\citenamefont {Fomin}\ \emph {et~al.}(2012)\citenamefont {Fomin}
  \emph {et~al.}}]{Fomin:2011ng}%
  \BibitemOpen
  \bibfield  {author} {\bibinfo {author} {\bibfnamefont {N.}~\bibnamefont
  {Fomin}} \emph {et~al.},\ }\href {\doibase 10.1103/PhysRevLett.108.092502}
  {\bibfield  {journal} {\bibinfo  {journal} {Phys. Rev. Lett.}\ }\textbf
  {\bibinfo {volume} {108}},\ \bibinfo {pages} {092502} (\bibinfo {year}
  {2012})}\BibitemShut {NoStop}%
\bibitem [{\citenamefont {Hen}\ \emph {et~al.}(2012)\citenamefont {Hen},
  \citenamefont {Piasetzky},\ and\ \citenamefont {Weinstein}}]{Hen:2012fm}%
  \BibitemOpen
  \bibfield  {author} {\bibinfo {author} {\bibfnamefont {O.}~\bibnamefont
  {Hen}}, \bibinfo {author} {\bibfnamefont {E.}~\bibnamefont {Piasetzky}}, \
  and\ \bibinfo {author} {\bibfnamefont {L.~B.}\ \bibnamefont {Weinstein}},\
  }\href {\doibase 10.1103/PhysRevC.85.047301} {\bibfield  {journal} {\bibinfo
  {journal} {Phys. Rev. C}\ }\textbf {\bibinfo {volume} {85}},\ \bibinfo
  {pages} {047301} (\bibinfo {year} {2012})}\BibitemShut {NoStop}%
\bibitem [{\citenamefont {Arrington}\ \emph
  {et~al.}(2012{\natexlab{a}})\citenamefont {Arrington} \emph
  {et~al.}}]{Arrington:2012ax}%
  \BibitemOpen
  \bibfield  {author} {\bibinfo {author} {\bibfnamefont {J.}~\bibnamefont
  {Arrington}} \emph {et~al.},\ }\href {\doibase 10.1103/PhysRevC.86.065204}
  {\bibfield  {journal} {\bibinfo  {journal} {Phys. Rev. C}\ }\textbf {\bibinfo
  {volume} {86}},\ \bibinfo {pages} {065204} (\bibinfo {year}
  {2012}{\natexlab{a}})}\BibitemShut {NoStop}%
\bibitem [{\citenamefont {Weinstein}\ \emph {et~al.}(2011)\citenamefont
  {Weinstein}, \citenamefont {Piasetzky}, \citenamefont {Higinbotham},
  \citenamefont {Gomez}, \citenamefont {Hen},\ and\ \citenamefont
  {Shneor}}]{Weinstein:2010rt}%
  \BibitemOpen
  \bibfield  {author} {\bibinfo {author} {\bibfnamefont {L.~B.}\ \bibnamefont
  {Weinstein}}, \bibinfo {author} {\bibfnamefont {E.}~\bibnamefont
  {Piasetzky}}, \bibinfo {author} {\bibfnamefont {D.~W.}\ \bibnamefont
  {Higinbotham}}, \bibinfo {author} {\bibfnamefont {J.}~\bibnamefont {Gomez}},
  \bibinfo {author} {\bibfnamefont {O.}~\bibnamefont {Hen}}, \ and\ \bibinfo
  {author} {\bibfnamefont {R.}~\bibnamefont {Shneor}},\ }\href {\doibase
  10.1103/PhysRevLett.106.052301} {\bibfield  {journal} {\bibinfo  {journal}
  {Phys. Rev. Lett.}\ }\textbf {\bibinfo {volume} {106}},\ \bibinfo {pages}
  {052301} (\bibinfo {year} {2011})}\BibitemShut {NoStop}%
\bibitem [{\citenamefont {Schmookler}\ \emph {et~al.}(2019)\citenamefont
  {Schmookler} \emph {et~al.}}]{CLAS:2019vsb}%
  \BibitemOpen
  \bibfield  {author} {\bibinfo {author} {\bibfnamefont {B.}~\bibnamefont
  {Schmookler}} \emph {et~al.},\ }\href {\doibase 10.1038/s41586-019-0925-9}
  {\bibfield  {journal} {\bibinfo  {journal} {Nature}\ }\textbf {\bibinfo
  {volume} {566}},\ \bibinfo {pages} {354} (\bibinfo {year}
  {2019})}\BibitemShut {NoStop}%
\bibitem [{\citenamefont {Arrington}\ and\ \citenamefont
  {Fomin}(2019)}]{Arrington:2019wky}%
  \BibitemOpen
  \bibfield  {author} {\bibinfo {author} {\bibfnamefont {J.}~\bibnamefont
  {Arrington}}\ and\ \bibinfo {author} {\bibfnamefont {N.}~\bibnamefont
  {Fomin}},\ }\href {\doibase 10.1103/PhysRevLett.123.042501} {\bibfield
  {journal} {\bibinfo  {journal} {Phys. Rev. Lett.}\ }\textbf {\bibinfo
  {volume} {123}},\ \bibinfo {pages} {042501} (\bibinfo {year}
  {2019})}\BibitemShut {NoStop}%
\bibitem [{\citenamefont {Tang}\ \emph {et~al.}(2003)\citenamefont {Tang} \emph
  {et~al.}}]{Tang:2002ww}%
  \BibitemOpen
  \bibfield  {author} {\bibinfo {author} {\bibfnamefont {A.}~\bibnamefont
  {Tang}} \emph {et~al.},\ }\href {\doibase 10.1103/PhysRevLett.90.042301}
  {\bibfield  {journal} {\bibinfo  {journal} {Phys. Rev. Lett.}\ }\textbf
  {\bibinfo {volume} {90}},\ \bibinfo {pages} {042301} (\bibinfo {year}
  {2003})}\BibitemShut {NoStop}%
\bibitem [{\citenamefont {Subedi}\ \emph {et~al.}(2008)\citenamefont {Subedi}
  \emph {et~al.}}]{Subedi:2008zz}%
  \BibitemOpen
  \bibfield  {author} {\bibinfo {author} {\bibfnamefont {R.}~\bibnamefont
  {Subedi}} \emph {et~al.},\ }\href {\doibase 10.1126/science.1156675}
  {\bibfield  {journal} {\bibinfo  {journal} {Science}\ }\textbf {\bibinfo
  {volume} {320}},\ \bibinfo {pages} {1476} (\bibinfo {year}
  {2008})}\BibitemShut {NoStop}%
\bibitem [{\citenamefont {Arrington}\ \emph
  {et~al.}(2012{\natexlab{b}})\citenamefont {Arrington}, \citenamefont
  {Higinbotham}, \citenamefont {Rosner},\ and\ \citenamefont
  {Sargsian}}]{Arrington:2011ax}%
  \BibitemOpen
  \bibfield  {author} {\bibinfo {author} {\bibfnamefont {J.}~\bibnamefont
  {Arrington}}, \bibinfo {author} {\bibfnamefont {D.}~\bibnamefont
  {Higinbotham}}, \bibinfo {author} {\bibfnamefont {G.}~\bibnamefont {Rosner}},
  \ and\ \bibinfo {author} {\bibfnamefont {M.}~\bibnamefont {Sargsian}},\
  }\href {\doibase 10.1016/j.ppnp.2012.04.002} {\bibfield  {journal} {\bibinfo
  {journal} {Prog. Part. Nucl. Phys.}\ }\textbf {\bibinfo {volume} {67}},\
  \bibinfo {pages} {898} (\bibinfo {year} {2012}{\natexlab{b}})}\BibitemShut
  {NoStop}%
\bibitem [{\citenamefont {Fomin}\ \emph {et~al.}(2017)\citenamefont {Fomin},
  \citenamefont {Higinbotham}, \citenamefont {Sargsian},\ and\ \citenamefont
  {Solvignon}}]{fomin17}%
  \BibitemOpen
  \bibfield  {author} {\bibinfo {author} {\bibfnamefont {N.}~\bibnamefont
  {Fomin}}, \bibinfo {author} {\bibfnamefont {D.}~\bibnamefont {Higinbotham}},
  \bibinfo {author} {\bibfnamefont {M.}~\bibnamefont {Sargsian}}, \ and\
  \bibinfo {author} {\bibfnamefont {P.}~\bibnamefont {Solvignon}},\ }\href
  {\doibase 10.1146/annurev-nucl-102115-044939} {\bibfield  {journal} {\bibinfo
   {journal} {Ann. Rev. Nucl. Part. Sci.}\ }\textbf {\bibinfo {volume} {67}},\
  \bibinfo {pages} {129} (\bibinfo {year} {2017})}\BibitemShut {NoStop}%
\bibitem [{\citenamefont {Arrington}\ \emph {et~al.}(2022)\citenamefont
  {Arrington}, \citenamefont {Fomin},\ and\ \citenamefont
  {Schmidt}}]{Arrington:2022sov}%
  \BibitemOpen
  \bibfield  {author} {\bibinfo {author} {\bibfnamefont {J.}~\bibnamefont
  {Arrington}}, \bibinfo {author} {\bibfnamefont {N.}~\bibnamefont {Fomin}}, \
  and\ \bibinfo {author} {\bibfnamefont {A.}~\bibnamefont {Schmidt}},\
  }\href@noop {} {\bibfield  {journal} {\bibinfo  {journal} {Ann. Rev. Nucl.
  Part. Sci.}\ }\textbf {\bibinfo {volume} {72}},\ \bibinfo {pages} {307}
  (\bibinfo {year} {2022})}\BibitemShut {NoStop}%
\bibitem [{\citenamefont {Arrington}\ \emph {et~al.}(2010)\citenamefont
  {Arrington}, \citenamefont {Daniel}, \citenamefont {Fomin},\ and\
  \citenamefont {Gaskell}}]{E1210008}%
  \BibitemOpen
  \bibfield  {author} {\bibinfo {author} {\bibfnamefont {J.}~\bibnamefont
  {Arrington}}, \bibinfo {author} {\bibfnamefont {A.}~\bibnamefont {Daniel}},
  \bibinfo {author} {\bibfnamefont {N.}~\bibnamefont {Fomin}}, \ and\ \bibinfo
  {author} {\bibfnamefont {D.}~\bibnamefont {Gaskell}},\ }\href@noop {}
  {}\bibinfo {howpublished} {spokespersons, Jefferson lab experiment
  E12-10-008, \url{https://www.jlab.org/exp_prog/proposals/10/PR12-10-008.pdf}}
  (\bibinfo {year} {2010})\BibitemShut {NoStop}%
\bibitem [{\citenamefont {Arrington}\ \emph {et~al.}(2006)\citenamefont
  {Arrington}, \citenamefont {Day}, \citenamefont {Fomin},\ and\ \citenamefont
  {Solvignon}}]{E1206105}%
  \BibitemOpen
  \bibfield  {author} {\bibinfo {author} {\bibfnamefont {J.}~\bibnamefont
  {Arrington}}, \bibinfo {author} {\bibfnamefont {D.}~\bibnamefont {Day}},
  \bibinfo {author} {\bibfnamefont {N.}~\bibnamefont {Fomin}}, \ and\ \bibinfo
  {author} {\bibfnamefont {P.}~\bibnamefont {Solvignon}},\ }\href@noop {}
  {}\bibinfo {howpublished} {spokespersons, Jefferson lab experiment
  E12-06-105, \url{https://www.jlab.org/exp_prog/proposals/06/PR12-06-105.pdf}}
  (\bibinfo {year} {2006})\BibitemShut {NoStop}%
\bibitem [{\citenamefont {Mo}\ and\ \citenamefont {Tsai}(1969)}]{Mo:1968cg}%
  \BibitemOpen
  \bibfield  {author} {\bibinfo {author} {\bibfnamefont {L.~W.}\ \bibnamefont
  {Mo}}\ and\ \bibinfo {author} {\bibfnamefont {Y.-S.}\ \bibnamefont {Tsai}},\
  }\href {\doibase 10.1103/RevModPhys.41.205} {\bibfield  {journal} {\bibinfo
  {journal} {Rev. Mod. Phys.}\ }\textbf {\bibinfo {volume} {41}},\ \bibinfo
  {pages} {205} (\bibinfo {year} {1969})}\BibitemShut {NoStop}%
\bibitem [{\citenamefont {Tsai}(1971)}]{Tsai:1971qi}%
  \BibitemOpen
  \bibfield  {author} {\bibinfo {author} {\bibfnamefont {Y.~S.}\ \bibnamefont
  {Tsai}},\ }\href@noop {} {\  (\bibinfo {year} {1971})},\ \bibinfo {note}
  {{SLAC-PUB-848}}\BibitemShut {NoStop}%
\bibitem [{\citenamefont {Dasu}\ \emph {et~al.}(1994)\citenamefont {Dasu},
  \citenamefont {deBarbaro}, \citenamefont {Bodek}, \citenamefont {Harada},
  \citenamefont {Krasny} \emph {et~al.}}]{Dasu:1993vk}%
  \BibitemOpen
  \bibfield  {author} {\bibinfo {author} {\bibfnamefont {S.}~\bibnamefont
  {Dasu}}, \bibinfo {author} {\bibfnamefont {P.}~\bibnamefont {deBarbaro}},
  \bibinfo {author} {\bibfnamefont {A.}~\bibnamefont {Bodek}}, \bibinfo
  {author} {\bibfnamefont {H.}~\bibnamefont {Harada}}, \bibinfo {author}
  {\bibfnamefont {M.}~\bibnamefont {Krasny}},  \emph {et~al.},\ }\href
  {\doibase 10.1103/PhysRevD.49.5641} {\bibfield  {journal} {\bibinfo
  {journal} {Phys. Rev. D}\ }\textbf {\bibinfo {volume} {49}},\ \bibinfo
  {pages} {5641} (\bibinfo {year} {1994})}\BibitemShut {NoStop}%
\bibitem [{\citenamefont {Bosted}\ and\ \citenamefont
  {Mamyan}(2012)}]{Bosted:2012qc}%
  \BibitemOpen
  \bibfield  {author} {\bibinfo {author} {\bibfnamefont {P.}~\bibnamefont
  {Bosted}}\ and\ \bibinfo {author} {\bibfnamefont {V.}~\bibnamefont
  {Mamyan}},\ }\href@noop {} {\bibfield  {journal} {\bibinfo  {journal}
  {arXiv:1203.2262}\ } (\bibinfo {year} {2012})}\BibitemShut {NoStop}%
\bibitem [{\citenamefont {Maieron}\ \emph {et~al.}(2002)\citenamefont
  {Maieron}, \citenamefont {Donnelly},\ and\ \citenamefont
  {Sick}}]{Maieron:2001it}%
  \BibitemOpen
  \bibfield  {author} {\bibinfo {author} {\bibfnamefont {C.}~\bibnamefont
  {Maieron}}, \bibinfo {author} {\bibfnamefont {T.~W.}\ \bibnamefont
  {Donnelly}}, \ and\ \bibinfo {author} {\bibfnamefont {I.}~\bibnamefont
  {Sick}},\ }\href {\doibase 10.1103/PhysRevC.65.025502} {\bibfield  {journal}
  {\bibinfo  {journal} {Phys. Rev. C}\ }\textbf {\bibinfo {volume} {65}},\
  \bibinfo {pages} {025502} (\bibinfo {year} {2002})}\BibitemShut {NoStop}%
\bibitem [{\citenamefont {Bodek}\ \emph {et~al.}(1979)\citenamefont {Bodek}
  \emph {et~al.}}]{Bodek:1979rx}%
  \BibitemOpen
  \bibfield  {author} {\bibinfo {author} {\bibfnamefont {A.}~\bibnamefont
  {Bodek}} \emph {et~al.},\ }\href {\doibase 10.1103/PhysRevD.20.1471}
  {\bibfield  {journal} {\bibinfo  {journal} {Phys. Rev. D}\ }\textbf {\bibinfo
  {volume} {20}},\ \bibinfo {pages} {1471} (\bibinfo {year}
  {1979})}\BibitemShut {NoStop}%
\bibitem [{\citenamefont {Gomez}\ \emph {et~al.}(1994)\citenamefont {Gomez}
  \emph {et~al.}}]{Gomez:1993ri}%
  \BibitemOpen
  \bibfield  {author} {\bibinfo {author} {\bibfnamefont {J.}~\bibnamefont
  {Gomez}} \emph {et~al.},\ }\href {\doibase 10.1103/PhysRevD.49.4348}
  {\bibfield  {journal} {\bibinfo  {journal} {Phys. Rev. D}\ }\textbf {\bibinfo
  {volume} {49}},\ \bibinfo {pages} {4348} (\bibinfo {year}
  {1994})}\BibitemShut {NoStop}%
\bibitem [{\citenamefont {Arrington}\ \emph {et~al.}(2021)\citenamefont
  {Arrington} \emph {et~al.}}]{Arrington:2021vuu}%
  \BibitemOpen
  \bibfield  {author} {\bibinfo {author} {\bibfnamefont {J.}~\bibnamefont
  {Arrington}} \emph {et~al.},\ }\href {\doibase 10.1103/PhysRevC.104.065203}
  {\bibfield  {journal} {\bibinfo  {journal} {Phys. Rev. C}\ }\textbf {\bibinfo
  {volume} {104}},\ \bibinfo {pages} {065203} (\bibinfo {year}
  {2021})}\BibitemShut {NoStop}%
\bibitem [{\citenamefont {Christy}()}]{christy_priv}%
  \BibitemOpen
  \bibfield  {author} {\bibinfo {author} {\bibfnamefont {M.~E.}\ \bibnamefont
  {Christy}},\ }\href@noop {} {}\bibinfo {note} {Private
  communication}\BibitemShut {NoStop}%
\bibitem [{\citenamefont {Arrington}\ \emph
  {et~al.}(2012{\natexlab{c}})\citenamefont {Arrington}, \citenamefont
  {Rubin},\ and\ \citenamefont {Melnitchouk}}]{Arrington:2011qt}%
  \BibitemOpen
  \bibfield  {author} {\bibinfo {author} {\bibfnamefont {J.}~\bibnamefont
  {Arrington}}, \bibinfo {author} {\bibfnamefont {J.~G.}\ \bibnamefont
  {Rubin}}, \ and\ \bibinfo {author} {\bibfnamefont {W.}~\bibnamefont
  {Melnitchouk}},\ }\href {\doibase 10.1103/PhysRevLett.108.252001} {\bibfield
  {journal} {\bibinfo  {journal} {Phys. Rev. Lett.}\ }\textbf {\bibinfo
  {volume} {108}},\ \bibinfo {pages} {252001} (\bibinfo {year}
  {2012}{\natexlab{c}})}\BibitemShut {NoStop}%
\bibitem [{\citenamefont {Aste}\ and\ \citenamefont
  {Jourdan}(2004)}]{Aste:2004yz}%
  \BibitemOpen
  \bibfield  {author} {\bibinfo {author} {\bibfnamefont {A.}~\bibnamefont
  {Aste}}\ and\ \bibinfo {author} {\bibfnamefont {J.}~\bibnamefont {Jourdan}},\
  }\href {\doibase 10.1209/epl/i2004-10113-x} {\bibfield  {journal} {\bibinfo
  {journal} {EPL}\ }\textbf {\bibinfo {volume} {67}},\ \bibinfo {pages} {753}
  (\bibinfo {year} {2004})}\BibitemShut {NoStop}%
\bibitem [{\citenamefont {Wang}\ and\ \citenamefont
  {Chen}(2015)}]{Wang:2014rxa}%
  \BibitemOpen
  \bibfield  {author} {\bibinfo {author} {\bibfnamefont {R.}~\bibnamefont
  {Wang}}\ and\ \bibinfo {author} {\bibfnamefont {X.}~\bibnamefont {Chen}},\
  }\href {\doibase 10.1016/j.physletb.2015.02.059} {\bibfield  {journal}
  {\bibinfo  {journal} {Phys. Lett. B}\ }\textbf {\bibinfo {volume} {743}},\
  \bibinfo {pages} {267} (\bibinfo {year} {2015})}\BibitemShut {NoStop}%
\bibitem [{\citenamefont {Pieper}\ and\ \citenamefont
  {Wiringa}(2001)}]{Pieper:2001mp}%
  \BibitemOpen
  \bibfield  {author} {\bibinfo {author} {\bibfnamefont {S.~C.}\ \bibnamefont
  {Pieper}}\ and\ \bibinfo {author} {\bibfnamefont {R.~B.}\ \bibnamefont
  {Wiringa}},\ }\href {\doibase 10.1146/annurev.nucl.51.101701.132506}
  {\bibfield  {journal} {\bibinfo  {journal} {Ann. Rev. Nucl. Part. Sci.}\
  }\textbf {\bibinfo {volume} {51}},\ \bibinfo {pages} {53} (\bibinfo {year}
  {2001})}\BibitemShut {NoStop}%
\bibitem [{\citenamefont {Samanta}\ and\ \citenamefont
  {Adhikari}(2002)}]{Samanta:2002sp}%
  \BibitemOpen
  \bibfield  {author} {\bibinfo {author} {\bibfnamefont {C.}~\bibnamefont
  {Samanta}}\ and\ \bibinfo {author} {\bibfnamefont {S.}~\bibnamefont
  {Adhikari}},\ }\href {\doibase 10.1103/PhysRevC.65.037301} {\bibfield
  {journal} {\bibinfo  {journal} {Phys. Rev. C}\ }\textbf {\bibinfo {volume}
  {65}},\ \bibinfo {pages} {037301} (\bibinfo {year} {2002})}\BibitemShut
  {NoStop}%
\bibitem [{\citenamefont {Heyde}\ and\ \citenamefont
  {Heyde}(1999)}]{heyde1999basic}%
  \BibitemOpen
  \bibfield  {author} {\bibinfo {author} {\bibfnamefont {K.}~\bibnamefont
  {Heyde}}\ and\ \bibinfo {author} {\bibfnamefont {K.}~\bibnamefont {Heyde}},\
  }\href {https://books.google.com/books?id=Lb3vAAAAMAAJ} {\emph {\bibinfo
  {title} {Basic Ideas and Concepts in Nuclear Physics, An Introductory
  Approach}}},\ Fundamental and applied nuclear physics series\ (\bibinfo
  {publisher} {Taylor \& Francis},\ \bibinfo {year} {1999})\BibitemShut
  {NoStop}%
\bibitem [{\citenamefont {Zhusupov}\ \emph {et~al.}(2020)\citenamefont
  {Zhusupov}, \citenamefont {Zhaksybekova},\ and\ \citenamefont
  {Kabatayeva}}]{Zhusupov:2020nkn}%
  \BibitemOpen
  \bibfield  {author} {\bibinfo {author} {\bibfnamefont {M.~A.}\ \bibnamefont
  {Zhusupov}}, \bibinfo {author} {\bibfnamefont {K.~A.}\ \bibnamefont
  {Zhaksybekova}}, \ and\ \bibinfo {author} {\bibfnamefont {R.~S.}\
  \bibnamefont {Kabatayeva}},\ }\href {\doibase 10.3103/S1062873820100317}
  {\bibfield  {journal} {\bibinfo  {journal} {Bull. Russ. Acad. Sci. Phys.}\
  }\textbf {\bibinfo {volume} {84}},\ \bibinfo {pages} {1175} (\bibinfo {year}
  {2020})}\BibitemShut {NoStop}%
\bibitem [{\citenamefont {Itagaki}\ \emph {et~al.}(2022)\citenamefont
  {Itagaki}, \citenamefont {Naito},\ and\ \citenamefont
  {Hirata}}]{Itagaki:2021fpk}%
  \BibitemOpen
  \bibfield  {author} {\bibinfo {author} {\bibfnamefont {N.}~\bibnamefont
  {Itagaki}}, \bibinfo {author} {\bibfnamefont {T.}~\bibnamefont {Naito}}, \
  and\ \bibinfo {author} {\bibfnamefont {Y.}~\bibnamefont {Hirata}},\ }\href
  {\doibase 10.1103/PhysRevC.105.024304} {\bibfield  {journal} {\bibinfo
  {journal} {Phys. Rev. C}\ }\textbf {\bibinfo {volume} {105}},\ \bibinfo
  {pages} {024304} (\bibinfo {year} {2022})}\BibitemShut {NoStop}%
\end{thebibliography}%

\end{document}